\documentstyle[11pt]{article} 
\title{\bf Neutrino mass patterns, 
 $R$-parity violating supersymmetry and associated phenomenology}

\author{Biswarup Mukhopadhyaya\thanks{Talk delivered at the
Discussion Meeting on Neutrino Physics, Physical Research Laboratory,
Ahmedabad, February 2-4, 1999}\\ \em Mehta Research Institute,
Chhatnag Road, Jhusi, Allahabad - 211 019, India}

\begin{document}


\maketitle

\begin{abstract}
Motivated by the recent SuperKamiokande results on atmospheric neutrinos, 
we incorporate massive neutrinos, with large angle 
oscillation between the second and third generations, in a theory
with $R$-parity violating supersymmetry. The general features of such a 
theory are briefly reviewed. We emphasize its testability 
through the observation of comparable numbers of 
muons and taus, produced together with the W-boson, in decays
of the lightest neutralino. A distinctly measurable decay gap is another
remarkable feature of such a scenario.  

\end{abstract}

\vskip 1 true cm


\def\baselinestretch{1.8}

\section{Introduction}
Although various options beyond the standard model (SM) of electroweak
interactions are being investigated with great interest for quite some
time now, the standard model has faced practically {\it no} experimental
contradictions in terrestrial experiments so far. In this respect, the 
observed results on solar
and atmospheric neutrinos have a unique role to play, in the sense that
their confirmation will require the existence of neutrino masses and
mixing, and therefore will take one beyond the jurisdiction of the
standard model. It is thus quite natural 
that the apparent oscillation of  
the muon neutrinos to another species, inferred with far 
greater confidence than before from the recent data 
from the SuperKamiokande (SK) experiment \cite{1},
is being enthusiastically examined for traces of some kind non-standard 
physics answering to a neutrino mass pattern of the suggested type. 
There are, however, a very number of possibilities to explore, and
the credibility of any one of them will depend not only
on how well they explain the neutrino data but also on their other
testable consequences. In this regard, one must say that the recent
developments in neutrino physics have triggered a lot of incisive thinking
on other areas of particle phenomenology as well. Here we propose to
discuss some such phenomenological issues in the particular context
of supersymmetric theories.

Supersymmetry (SUSY) is perhaps the object of the hottest 
pursuit in terms of physics beyond the SM \cite{2}. 
Its usefulness in solving the naturalness problem,
its tantalisingly spectacular role in achieving the unification of
coupling constants, and its almost invariable presence in theories
attempting to unify gravity with the other interactions make it an
extremely appealing theoretical option. However, there is no concerete
experimental evidence in its favour yet. It is therefore 
quite natural that
the possibilities of generating neutrino masses and 
mixing in a SUSY scenario
should be investigated, especially when evidences for the latter are
already knocking at our doors. 

Neutrino masses will either necessitate the existence of 
right-handed neutrinos or require  violation of lepton number (L) so that
Majorana masses are possible. The former possibility entails an
augmentation of the particle content of the minimal SUSY 
standard model (MSSM).
The latter one does not require it, 
but forces one to go beyond the minimal 
model again, whereby lepton number violation can be 
allowed in the theory. 
However, such a violation is inbuilt in those SUSY 
theories where R-parity, defined 
as $R = (-1)^{3B + L + 2S}$, is not a conserved quantity 
anymore \cite{3}. 
This is quite consistent with the absence of proton decay so 
long as baryon number (B)
is not violated simultaneously, a situation that again may arise in SUSY
where there are scalar leptons and baryons and therefore L-violation and
B-conservation does not interfere with the gauge current structure of the
theory. 

In the next section we present a summary on R-parity violating models,
with an emphasis on the type which has a key role
in our claims, namely, one with R-parity violation through bilinear terms.
In the same section we also discuss the generation of neutrino masses 
in such theories both at the tree-and one-loop levels. Some distinct
accelerator signals, of one viable scenario at least, are mentioned in
section 3. We conclude in section 4.

\section{R-parity violation and neutrino mass}
The MSSM superpotential is given by
\begin{equation}
W_{MSSM} = {\mu} {\hat H}_1 {\hat H}_2 + h_{ij}^l {\hat L}_i {\hat
H}_1 {\hat E}_j^c
+ h_{ij}^d {\hat Q}_i {\hat H}_1 {\hat D}_j^c + h_{ij}^u {\hat Q}_i
{\hat H}_2 {\hat U}_j^c
\end{equation}
where the last three terms give the Yukawa interactions corresponding to
the masses of the charged leptons and the down-and up-type quarks, and
$\mu$ is the Higgsino mas parameter.

When R-parity is violated, the following additional terms can be added to
the superpotential: 
\begin{equation}
W_{\not R} = \lambda_{ijk} {\hat L}_i {\hat L}_j {\hat E}_k^c +
\lambda_{ijk}' {\hat L}_i {\hat Q}_j {\hat D}_k^c +
\lambda_{ijk}''{\hat U}_i^c {\hat D}_j^c {\hat D}_k^c + \epsilon_i {\hat
L}_i {\hat H}_2
\end{equation}
with the $\lambda''$-terms causing B-violation, and the remaining
ones, L-violation. In order to suppress proton decay, it is customary
(though not essential) to have one of the two types of 
nonconservation at a time. In the rest of this article, we will consider
only lepton numer violating effects.

The $\lambda$-and $\lambda'$-terms have been 
widely studied in conection 
with phenomenological consequences, enabling one to impose various 
kinds of limits on them \cite{4}. Their contributions to 
neutrino masses can be only through loops \cite{5}, and 
their multitude (there are 36 such couplings
altogether) makes the necessary adjustments possible for reproducing
the requisite values of neutrino masses and mixing angles. We shall come 
back to these `trilinear' effects later.

More interesting, however, are the three bilinear terms 
$\epsilon_{i}L_{i}H_2$ \cite{6}. There being only three terms of this 
type, the model looks simpler and more predictive with them alone as
sources of R-parity violation. This is particularly so because
the physical effects of the trilinear terms can be generated from the
bilinears by going to the appropriate bases. In addition, they  have
interesting consequences of their own \cite{7,8}, since terms of the type 
$\epsilon_{i}L_{i}H_2$ imply mixing between the Higgsinos and the 
charged leptons and neutrinos. In  this discusion, we shall assume,
without ay loss of generality,
the existence of such terms involving onl the second and third 
famililies of leptons.  

In the above scenario, the scalar potential contains the following
terms which are bilinear in the scalar fields:
\begin{eqnarray}
V_{\rm scal} &=& m^2_{L_3} {\tilde L}^2_3 + m^2_{L_2} {\tilde L}^2_2 + 
m^2_1 H^2_1 + m^2_2 H^2_2 + B \mu H_1 H_2 \nonumber\\*
&& + B_2 \epsilon_2 {\tilde L}_2 H_2 +B_3 \epsilon_3 {\tilde L}_3 H_2 
+ \mu \epsilon_3 {\tilde L}_3 H_1 + \mu \epsilon_2 {\tilde L}_2 H_1 + .....
\end{eqnarray} 
where $m_{L_i}$  denotes the mass of the {\it i}th scalar doublet
at the electroweak scale, and $m_1$ and $m_2$ are the mass parameters
corresponding to the two Higgs doublets. $B$, $B_2$ and $B_3$ are  
soft SUSY-breaking parameters.  

An immediate consequence of the additional (L-violating) soft terms in
the potential is a set of non-vanishing vacuum expectation values
(vev) for the sneutrinos \cite{9}. 
This gives rise to the mixing of the gauginos with neutrinos (and
charged leptons) through the sneutrino-neutrino-neutralino (and
sneutrino-charged lepton-chargino) interaction
terms.

By virtue of both the types of mixing described above, the hitherto
massless neutrino states enter into the neutralino mass matrix. This
leads to see-saw masses acquired by them via mixing with 
massive states. The parameters controlling the neutrino sector in
particular and R-parity violating effects in general are the 
bilinear coefficients $\epsilon_2$ , $\epsilon_3$  and the  
soft parameters $B_2$, $B_3$. For our purpose, however, it
is more convenient to eliminate the latter in favour of the sneutrno
vev's using the conditions of electroweak symmetry breaking \cite{7}.

For a better understanding, let us perform a basis rotation \cite{10},
removing the R-parity violating bilinear terms via a redefinition of
the lepton and Higgs superfields. This, however, does not eliminate
the effects of the bilinear terms, since they now take refuge in the
scalar potential. The sneutrino vev's in this rotated basis (which are
functions of both and the $\epsilon$'s and the soft terms in the
original basis) are instrumental in triggering neutrino-neutralino
mixing. Consequently, the $6 \times 6$ neutralino mass matrix in this
basis has the following form:
\begin{equation}
{\cal M} =  \left( \begin{array}{cccccc}
  0 & -\mu & \frac {gv} {\sqrt{2}} & 
  -\frac {g'v} {\sqrt{2}} & 0 & 0 \\
  -\mu & 0 & -\frac {gv'} {\sqrt{2}} 
       & \frac {g'v'} {\sqrt{2}} & 0 & 0 \\
 \frac {gv} {\sqrt{2}} & -\frac {gv'} {\sqrt{2}} & M & 0 & -\frac {gv_3} 
 {\sqrt{2}} & -\frac {gv_2} {\sqrt{2}} \\
 -\frac {g'v} {\sqrt{2}} & \frac {g'v'} {\sqrt{2}} & 0 & M' & 
  \frac {g'v_3} {\sqrt {2}} & \frac {g'v_2} {\sqrt {2}} \\
 0 & 0 & -\frac {gv_3} {\sqrt {2}} & \frac {g'v_3} {\sqrt {2}} & 
 0 & 0  \\
 0 & 0 & -\frac {gv_2} {\sqrt {2}} & \frac {g'v_2} {\sqrt {2}} & 
 0 & 0 
 \end{array}  \right)    
\end{equation}                                 
where the successive rows and columns correspond to
(${\tilde H}_2, {\tilde H}_1, -i\tilde{W_3}, 
-i\tilde{B}, \nu_\tau, \nu_\mu$),   $\nu_\tau$ and  $\nu_\mu$ being the
neutrino flavour eigenstates in this basis. Also,  with the sneutrino
vev's denoted by $v_2$ and $v_3$, 
$$
v\ \ (v') = \sqrt{2}\ {\left(\frac {m^2_Z} {\bar{g}^2} 
 - \frac {v^2_2+v_3^2} {2} \right)}
^{\frac {1} {2}} {{\sin} \beta}\ \ ({{\cos} \beta})
$$
$M$ and $M'$ being the ${\rm SU(2)}$ and ${\rm U(1)}$ gaugino mass
parameters respectively, and $\bar{g}=\sqrt{g^2+{g'}^2}.$

Next, one can define two states $\nu_3$ and $\nu_2$, where
\begin{equation}
\nu_3 =\cos \theta\ \nu_\tau  + \sin \theta\ \nu_\mu 
\end{equation}
and $\nu_2$ is the orthogonal combination, the neutrino mixing angle
being given by
\begin{equation}
\cos \theta = \frac{v_3}{\sqrt{v^2_2 + v^2_3}}
\end{equation}                  
Clearly, the state $\nu_3$ --- which alone develops cross-terms with
the massive gaugino states --- develops a see-saw type mass at the
tree-level. The orthogonal combination $\nu_2$ still remains massless.

An approximate expression (neglecting higher order terms in 
${m_z}/\mu$) for the tree-level neutrino mass is 
\begin{equation}
m_{\nu_3}\approx 
-\frac{\bar{g}^2 (v_2^2+v_3^2)}{2\ \bar{M}}\times
\frac{\bar{M}^2}{M M' - m_Z^2\ \bar{M}/\mu \ \sin 2\beta }
\end{equation}
where $\bar{g}^2 \bar{M}=g^2 M'+{g'}^2 M.$ The first term is very
similar to the usual see-saw formula, with the only difference that
couplings between the light and the heavy states is in the present
case due to gauge interactions.

The massive state $\nu_3$ can be naturally used to account for
atmospheric neutrino oscillations, with $\Delta m^2 = m_{\nu_3}^2.$
Large angle mixing between the $\nu_\mu$ and the $\nu_\tau$
corresponds to the situation where $v_2 \simeq v_3$.

The tree-level mass here is clearly controlled by the quantity $v' =
\sqrt{v_2^2 + v_3^2}$. This quantity, defined as the `effective'
sneutrino vev in the basis where the $\epsilon$'s are rotated away,
can be treated as a basis-independent measure of R-parity violation in
such theories. The SK data on atmospheric neutrinos restrict $v'$ to
be on the order of a few hundred keV's \cite{11}.  However, it should
be remembered that $v'$ is a function of $\epsilon_2$ and $\epsilon_3$
both of which can still be as large as on the order of the electroweak
scale. It has, for example, been shown \cite{12} that in models based
on N=1 supergravity, it is possible to have a very small value of $v'$
starting from large $\epsilon$'s, provided that one assumes the
R-conserving and R-violating soft terms to be of the same order at the
scale of dynamical SUSY breaking at a high energy.

Also, one has to address the question as to whether the treatment of
$\nu_3$ and $\nu_2$ as  mass eigenstates is proper,
from the viewpoint of the charged lepton mass marix being diagonal
in the basis used above. In fact, it can be shown that this is strictly
possible when $\epsilon_2$ is much smaller than $\epsilon_3$, failing
which one has to give a further basis rotation to defne the
neutrino mass eigenstates. However, the observable consequences
that we describe in the following section are found to be equally
valid, with the requirement shifted from the angle $\theta$ to the 
effective mixing angle to be in the neighbourhood of maximality.

Furthermore, a close examination of the scalar potential in such a
scenario reveals the possibility of additional mixing among the
charged sleptons, whereby flavour-changing neutral currents (FCNC) can
be enhanced. It has been concluded after a detailed study \cite{13}
that the supression of FCNC requires one to have the
$\epsilon$-parameters to be small compared to the MSSM parameter $\mu$
(or, in other words, to the electroweak scale) unless there is a
hierarchy between $\epsilon_2$ and $\epsilon_3$.

However, one still needs to find a mechanism for mass-splitting
between the massless state $\nu_2$ and the electron neutrino, and
to explain the solar neutrino puzzle \cite{14}. This is found to follow
naturally if one allows for R-parity (L) violating terms of all types
in the superpotential. The existence of the various 
$\lambda$ and $\lambda'$-terms will give rise to
loop conributions to the neutrino mass matrix. The generic expression
for such loop-induced masses is
\begin{equation}
(m^{loop}_\nu)_{ij} \simeq \frac {3} {8\pi^2}  m^d_k m^d_p M_{SUSY} 
\frac {1} {m^2_{\tilde q}} {\lambda_{ikp}'\lambda_{jpk}'}  + 
\frac {1} {8\pi^2}  m^l_k m^l_p M_{SUSY} 
\frac {1} {m^2_{\tilde l}} {\lambda_{ikp}\lambda_{jpk}}  
\end{equation}         
where $m^{d,(l)}$ denote the down-type quark (charged lepton) masses.
${m^2_{\tilde l}}$, ${m^2_{\tilde q}}$ are the slepton and 
squark mass squared. $M_{SUSY}(\sim \mu)$ is the effective scale of 
supersymmetry breaking. The mass eigenvalues
can be obtained by including the above loop contributions
in the mass matrix.

If we want the mass thus induced for the second generation neutrino to
be the right one to solve the solar neutrino problem, then one obtains
some constraint on the value of the $\lambda'$s as well as $\lambda$s.
In order to generate a splitting between the two residual massless
neutrinos, $\delta m^2 \simeq 5 \times 10^{-6}\ {\rm {eV^2}}$ (which
is suggested for an MSW solution ), a SUSY breaking mass of about 500
GeV implies $\lambda'\ (\lambda) \sim 10^{-4}~-~10^{-5}$.  The
mass-squared difference required for a vacuum oscillation solution to
the solar puzzle requires even smaller values of $\lambda'(\lambda)$.

\section{Phenomenological consequences}
As we have observed before, the SK data imply a constraint on the
basis-independent parameter $v'$. The allowed range of neutrino
mass-squared difference from the SK data, combining the fully
contained events, partially contained events and upward-going muons,
is about $1.5~-~6.0 \times 10^{-3}~eV^2$ at 90\% C.L. \cite{15}.  For
the lightest neutralino mass varying between 50 and 200 GeV, this
constrains $v'$ to be in the approximate range $0.0001~-~0.0003~ GeV$.

The experimentally observed signals characteristic of the scenario
described above should naturally be associated with decays of the
lightest nutralino, since that is a process where contributions
from R-parity volating effects will not face any competitions from
MSSM processes. 

In presence of only the trilinear R-violating terms in the
superpotential, the lightest neutralino can have various three-body
decay modes which can be genericaly described by $\chi^{0}
\longrightarrow \nu f \bar{f}$ and $\chi^{0} \longrightarrow l f_1
\bar{f_2}$, $f$, $f_1$ and $f_2$ being different quark and lepton
flavours that are kinematically allowed in the final state.

We have already seen that an important consequence of the bilinears is
a mixing between neutrinos and neutralinos as also between charged
leptons and charginos. This opens up additional decay chanels for the
lightest neutralino, namely, $\chi^{0} \longrightarrow l W$ and
$\chi^{0} \longrightarrow \nu Z$.  When the neutralino is heavier than
at least the W, these two-body channels dominate over the three-body
ones over a large region of the parameter space, the effect of which
can be observed in colliders such as the upgraded Tevatron, the LHC
and the projected high-energy electron-positon collider. Different
observables related to these decays have been studied in recent times
\cite{16}.

Here we would like to stress upon one distinctive feature of the
scenario that purportedly explains the SK results with the help of
bilinear R-parity violating terms. It has been found that over almost
the entire allowed range of the parameter space in this connection,
the lightest neutralino is dominated by th Bino.  A glance at the
neutralino mass matrix reveals that decays of the neutralino ($\simeq$
Bino) in such a case should be determined by the coupling of different
candidate fermionic fields in the final state with the massive
neutrino field $\nu_3$ which has a cross-term with the Bino. Large
angle neutrino mixing, on the the hand, implies that $\nu_3$ should
have comparable strengths of coupling with the muon and the tau. Thus,
a necessary consequence of the above type of explanation of the SK
results should be comparable numbers of muons and tau's emerging from
decays of the lightest neutralino, together with a $W$-boson in each
case \cite{10}.  Such signals, particlarly those in the form of muons
from two-body decays of the lightest neutralino, should distinguish
such a scenario. For further details including plots of the branching
ratios, the reader is referred to [10].

Of course, the event rates in the channel mentioned aboe will depend on
whether the two-body decays mentioned above indeed dominate over the
three-body decays. The latter are controlled by the size of the
$\lambda$-and $\lambda'$-parameters. It has been found that
if in this case these parameters have to be of the right orders of 
magnitude to explain the mass-splitting required by the solar neutrino
deficit, then, even for the MSW case, the decay widths driven by
the trilinear term are smaller than thsoe for the two-body decays by at 
least an order of magnitude. For vacuum oscillation, the three body 
decays turn out to be even smaller. Thus the prediction of comparable
numbers of muons and tau's seem to be quite robust so long as the 
two-body neutralino decays are kinematically allowed. 

The other important consequence \cite{10} of this picture is a large
decay length for the lightest neutralino. We have already mentioned
that the atmospheric neutrino results restrict the basis-independent
R-violating parameter $v'$ to the rather small value of a few hundred
keV's. This value affects the mixing angle involved in calculating the
decay width of the neutralino, which in turn is given by the formula
\begin{equation}
L =  \frac{\hbar}{\Gamma} 
\times 
\frac{p}{M(\tilde{\chi}^0_1)} 
\end{equation}  
where $\Gamma$ is the decay width of the lightest neutralino and $p$,
its momentum.  As can be seen from figure 2 in reference [10], the
decay length decreases for higher neutrino masses, as a result of the
enhanced probability of the flip between the Bino and a neutrino, when
the LSP is dominated by the Bino. Also, a relatively massive
neutralino decays faster and hence has a smaller decay length. The
interesting fact here is that even for a neutralino as massive as 250
GeV, the decay length is as large as about 0.1 to 10 millimeters.
This clearly will leave a measurable decay gap, which unmistakably
characterises the theoretical construction under investigation here
\cite{17}.

If the lightest neutralino can have two-body charged current decays,  
then the Majorana character of the latter also leads to the possibility 
of like-sign dimuons and ditaus from pair-produced neutralinos. Modulo
the efficiency of simultaneous identifiation of W-pairs, these
like-sign dileptons can also be quite useful in verifying the type
of theory discussed here.

\section{Summary and conclusions}
We have demonstrated that it is posible to explain both the atmospheric
and solar neutrino deficits in a SUSY model with R-parity violation
inbuilt in it. An important role is played by the blinear R-violating terms
in the superpotential, whereby a tree-level mass for one neutrino can
be generated via mixing with neutralinos. The mass-squared difference expected
from the atmospheric muon neutrino deficiency (for $\nu_\mu - \nu_\tau$
oscillation) constrains the basis-independent parameter characterising
R-parity violation in the neutrino-neutralino and lepton-chargino sectors.
Side by side, the existence of trilinear lepton number violating 
terms in the superpotential can give rise to a mass-splitting between
the two remaining neutrinos and thus account for the 
solar neutrino deficit.
The values of the trilinear parameters required for this imply that
the lightest neutralino should dominantly decay in two-body channels
if it is heavier than the W-boson. Maximal mixing, as required by the
SuperKamiokande data, implies that comparable numbers of muons and tau's
should be seen in charged current decays of the neutralino when the
two-body decays are kinematically allowed. In addition, the magnitudes 
of the R-parity violating parameters required by the atmospheric neutrino
data causes the neutralinos to have large decay lengths, and therefore
leads to displaced vertices in SUSY search experiments. 
Thus R-parity violating 
SUSY lends itself as a viable mechanism for  generating the expected
neutrino masses and mixing patterns, with verifiable (or falsifiable)
consequences in collider experiments.

\paragraph*{Acknowledgements:} I am grateful to my collaborators,
Asesh K. Datta, Sourov Roy and Francesco Vissani, from whom I have 
learnt most of what has been discussed above. I also acknowledge useful
discussions with Anjan Joshipura. Finally, I thank the organisers of the 
Discussion Meeting on Neutrino Physics at the Physical Research Laboatory,
Ahmedabad, for giving me an opportunity to present my views on this
subject.

\end{document}